\documentclass[12pt,preprint]{aastex}

\begin{document}

\slugcomment{Accepted to ApJ Letters}

\title{The Algol triple system spatially resolved at optical wavelengths}

\author{R.~T. Zavala\altaffilmark{1}, C.~A. Hummel\altaffilmark{2}, 
D.~A. Boboltz\altaffilmark{3}, R. Ojha\altaffilmark{3,4}, 
D.~B. Shaffer\altaffilmark{5}, C. Tycner\altaffilmark{6}, 
M.~T. Richards\altaffilmark{7}, D.~J. Hutter\altaffilmark{1}}

\altaffiltext{1}{U.S. Naval Observatory, Flagstaff Station, 10391 W. Naval
Obs. Rd., Flagstaff, AZ 86001\email{bzavala,djh@nofs.navy.mil}}
\altaffiltext{2}{European Organization for Astronomical Research in the Southern 
Hemisphere, Karl-Schwarzschild-Str. 2, 85748 Garching bei M\"{u}nchen, 
Germany\email{chummel@eso.org}}
\altaffiltext{3}{U.S. Naval Observatory, 3450 Massachusetts Ave. NW, Washington D.C. 
20392\email{dboboltz,rojha@usno.navy.mil}}
\altaffiltext{4}{NVI, Inc., 7257 Hanover Parkway, Greenbelt, MD 20770}
\altaffiltext{5}{Lowell Observatory, 1400 W. Mars Hill Rd., Flagstaff, 
AZ 86001\email{shaffer@alumni.caltech.edu}}
\altaffiltext{6}{Department of Physics, Central Michigan University, Mt. Pleasant, 
MI 48859\email{c.tycner@cmich.edu}}
\altaffiltext{7}{Department of Astronomy and Astrophysics, Pennsylvania State University, 
525 Davey Lab, University Park, PA 19104\email{mrichards@astro.psu.edu}}

\begin{abstract}
Interacting binaries typically have separations in the milli-arcsecond 
regime and hence it has been challenging to resolve them at any wavelength.  
However, recent advances in optical interferometry have improved our 
ability to discern the components in these systems and have now enabled 
the direct determination of physical parameters. We used the Navy 
Prototype Optical Interferometer to produce for the first time
images resolving all three components in the well-known Algol triple system.   
Specifically, we have separated the tertiary component from the binary 
and simultaneously resolved the eclipsing binary pair, which represents 
the nearest and brightest eclipsing binary in the sky. We present 
revised orbital elements for the triple system, and we have rectified 
the 180-degree ambiguity in the position angle of Algol C. Our 
directly determined magnitude differences and masses for this 
triple star system are consistent with earlier light curve modeling results.  
\end{abstract}


\section{Introduction}

Algol ($\beta$ Per, HR 936), the prototype for a well known class of eclipsing 
binaries, has well over 200 years of published observations available for study. 
The eclipsing nature of the system was first suggested by \citet{good} when he stated 
the light variations could result from ``... the interposition of a large 
body revolving around Algol ...''. A long suspected third component in Algol was 
spectroscopically confirmed by \citet{ss57} and \citet{ebb}. \citet{soder} provides a 
review of Algol which is a useful starting 
point for summarizing this triple system. We will refer to the 
B-type primary star in Algol as Algol A, the K type secondary star as Algol 
B, and the more distant Am companion as Algol C \citep[see][for the 
exact spectral types]{mtr93}. It is thus a 
hierarchical triple system as defined by \citet{evans}. 
\citet{wh72} made the first radio detection of Algol and 
\citet{les93} determined that the radio emission 
in Algol comes from Algol B using multi-epoch Very Long Baseline Interferometry
observations. Algol was one of 12 radio stars used to link the HIPPARCOS optical 
reference frame to the International Celestial Reference System (ICRS) \citep{kov}. 
The 1.86 year orbit of Algol C was resolved by speckle interferometry \citep{lab74} 
and optical interferometry \citep{pan} but these investigations could not resolve 
the close binary. Also, these results suffer from a 180\arcdeg\ ambiguity in the 
absolute position angle (PA) due to a lack of phase information \citep{lab70}. 
The Fourth Catalog of Interferometric Measurements of Binary 
Stars has a more complete listing of these 
observations\footnote{http://www.usno.navy.mil/USNO/astrometry/ 
optical-IR-prod/wds/int4 \citep{int}}.

This considerable body of knowledge still leaves room for additional 
exploration, and permitted inconsistent descriptions 
of the orbital elements of the triple system. It would appear that 
the orbital elements listed in \citet{soder} and \citet{pan} 
would adequately describe the Algol triple system. But these 
elements include a 180\arcdeg\ difference in the position angle 
of the ascending node when compared with the orbital solution 
in the Hipparcos catalog \citep{linde}. Algol's role as one of a 
small number of radio stars used to link the HIPPARCOS optical 
reference frame to the ICRF requires resolution of this inconsistency.
Algol can also serve as a position angle calibrator for optical and 
near-IR interferometers and this strengthens the case for a resolution 
of this disagreement.   

Recently, Algol was observed in 
the near IR (K$_{s}$ = 2.133 $\mu$m) with the CHARA array and at 5 GHz with the 
European VLBI Network during December 2006 \citep{csiz}. The CHARA array 
with approximately 200 m baselines has a similar spatial resolution 
to the NPOI at optical wavelengths with 64 m baselines. The CHARA 
data resolve the close binary but \citet{csiz} make no mention 
of detecting Algol C. Csizmadia et al. use their VLBI and CHARA 
array observations to produce an orbit of Algol A$-$B with an opposite 
sense of rotation from that determined by \citet{les93}. We are then 
presented with another inconsistency in published results for the Algol system. 

During 2006 October and November we collected observations of Algol using the 
Navy Prototype Optical Interferometer (NPOI), the Very Long Baseline Array 
(VLBA)\footnote {The National Radio Astronomy 
Observatory is operated by Associated Universities, Inc., under cooperative agreement 
with the National Science Foundation.}, and the Lowell Observatory 42'' Hall 
telescope equipped with the solar$-$stellar spectrograph. We defer discussion 
of the radio and spectroscopic results for a future paper. The primary aim of this 
project was to resolve the close pair in Algol with the NPOI, and perform an 
absolute astrometric registration of the optical NPOI images to the 
ICRF phase-referenced VLBA images. As the NPOI records visibility squared 
and closure phase data we can determine the position angle calibration 
without a 180\arcdeg\ ambiguity and resolve the inconsistency between the 
orbital elements in \citet{soder} and \citet{pan} and the orbital 
elements in the HIPPARCOS catalog. We can also address the inconsistency 
in the direction of the close binary orbit between \citet{les93} and 
\citet{csiz}. 



In this paper, we report on our NPOI observations which extend the direct 
knowledge of the Algol triple system to optical wavelengths. 
In \S2 we discuss our observations, with particular attention to the 
calibration of the NPOI absolute position angles. In \S3 we present 
the astrometric orbits of the Algol A$-$B and AB$-$C systems. 
We conclude with \S4 and a comparison with light curve solutions and a 
discussion of the astrometric results. 

\section{Observations and Calibration}

Algol was observed with the NPOI from 1997 October to 2008 October 
(see Table~\ref{tbl-1}).  
The NPOI is a six element optical interferometer, 
described in detail in \citet{arm98}. Details regarding the general 
NPOI observational setup and data recording can be found in \citet{hum03} 
and \citet{ben03}. We combined the astrometric siderostats on the 
center (AC), east (AE), and west (AW) stations with the E6 and W7 imaging 
siderostats. The addition of the latter allowed projected baselines of up 
to 64 m in length, separated by about 60 degrees in position angle. 
Up to three baselines were recorded on each of the two spectrometers. 
We switched between two different four-station configurations 
half-way through the night. Post-processing of the data was performed 
using C. Hummel's OYSTER software package.  
The calibrator star for the 2006 observations was $\epsilon$ Persei 
(HR 1220, V=2.89, B0.5V, parallax ($\pi$) = 6.06 mas), 
located 9.6\arcdeg\ from Algol and an estimated diameter (d$_{est}$) 
of 0.43 mas based on its $R - I$ color \citep{moz,nat}. The 
2008 observations included $\epsilon$ Per, check binary $\gamma$ Per 
(HR 915), and an additional calibrator for $\gamma$ Per, $\epsilon$ 
Cassiopeiae (HR 542, V=3.34, B3III, $\pi$ = 7.38 mas, d$_{est} = 0.43$ mas). 
The uncertainties on the estimated diameters are 10\%. Since $\gamma$ 
Per is 16.2\arcdeg\ from $\epsilon$ Per, Algol and $\gamma$ Per could share 
the same calibrator. $\epsilon$ Cas then serves as a secondary check on 
the calibration. 

To verify the absolute NPOI PA calibration 
using our 2008 Oct 27 observation of $\gamma$ Per we fit a model to the observed 
squared visibilities and triple phases in OYSTER and imaged $\gamma$ Per 
using DIFMAP \citep{dif1,dif2}. The expected position of $\gamma$ Per was 163.0 mas at 
244.6\arcdeg\ with an expected R band magnitude difference of 1.5$-$1.6 
\citep{pri03}. We observed  $\gamma$ Per at a position of 
160.92 $\pm$ 0.28 mas at 244.95 $\pm$ 0.22\arcdeg\ and a 
magnitude difference (800nm) of 1.50 $\pm$ 0.08. The same result was 
obtained for both choices ($\epsilon$ Per or $\epsilon$ Cas) of the 
calibrator stars which verifies the absolute PA calibration of the NPOI.

\section{Analysis and results}

The combined visibility data for each night allow the determination
of the relative positions of the components. Note that the orbital 
motion of the A$-$B pair is significant during each night's observation. 
Reliable estimates of the magnitude differences were available in the 
literature and were used as an initial guess in our model fits.  
The dominant feature of a finite magnitude difference
between the C component and the combined light
of the A$-$B pair in the data is 
the pronounced sinusoidal variation of the squared visibilities.
Superimposed on this variation is a subtle modulation due to the 
(larger) magnitude difference of the close binary itself. 
The magnitude differences in the $V$ ($\Delta V=2.92$) and 
Cousins-I ($\Delta I=2.63$) bands between the eclipsing pair and component C was 
determined previously by  \citet{pan} using the Mark III Stellar Interferometer.
An initial estimate for the magnitude difference between A and B was determined from 
the light curve analysis by \citet{mtr88} in the V band ($\Delta V=3.92$), and we 
estimated the value for the NPOI 800 nm filter ($\Delta I=2.6$) using the effective 
temperatures and log {\it g} values of \citet{mtr93} and the Kurucz model 
atmospheres \citep{kur}. 
As a check, the same procedure correctly reproduces the J-band magnitude
difference given by \citet{mtr88}. Since the diameters are only 
barely resolved we adopted the values given in \citet{mtr88} for 
components A, B, and C, converted to angular diameters using the parallax 
of Algol of 35.1 mas (distance $28.5$ pc, distance modulus $2.27$, 
\citet{esa97}). The stellar parameters initially used to model the Algol 
triple system are listed in Table~\ref{tbl-2} and  
were kept fixed for the fits of the relative component positions to refine the 
orbital elements. After these initial refinements of the orbital elements 
we also fit for the masses. 

\begin{figure}
\plotone{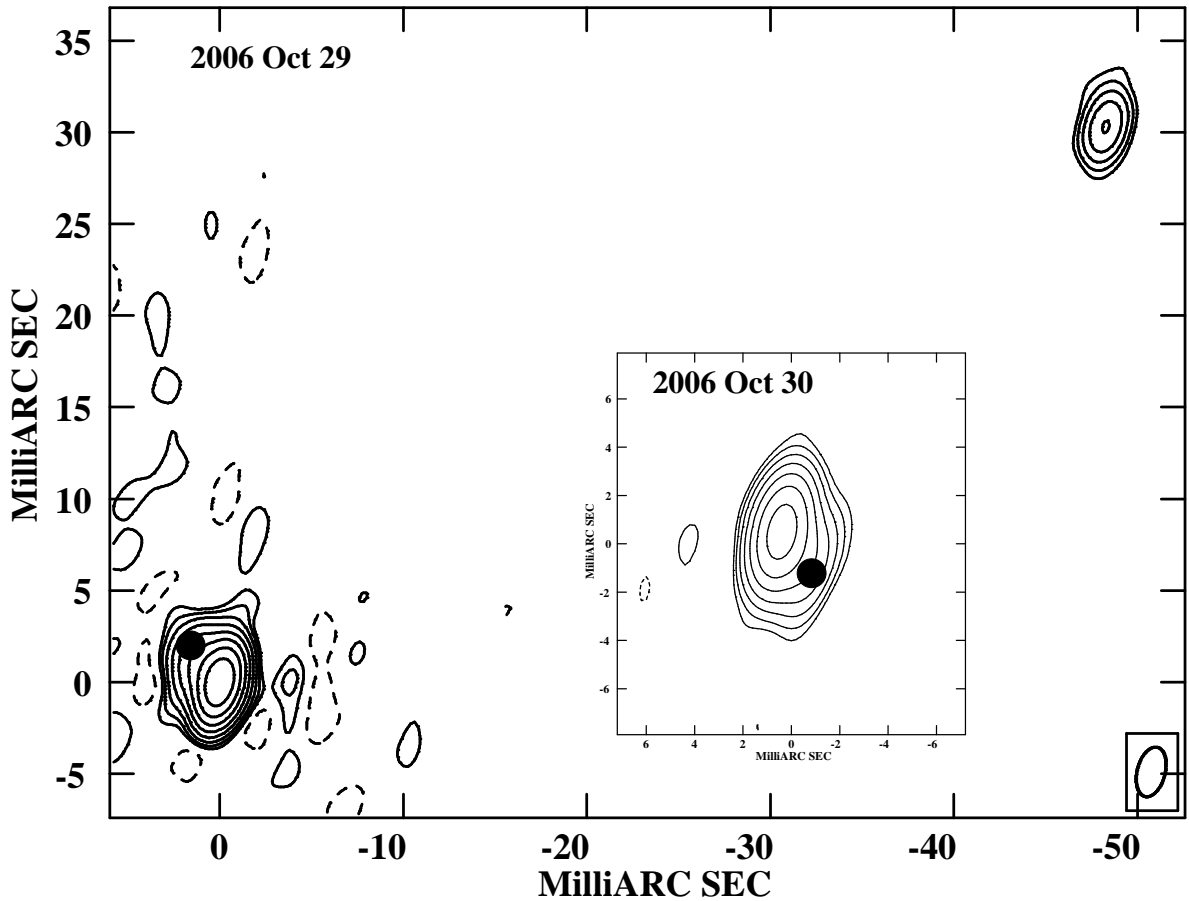}
\caption{An image of the Algol triple system made from the NPOI data 
of 2006 Oct 29. Algol C is the component in the upper right hand corner. 
The inset shows a close-up image made from the 
NPOI observation of 2006 Oct 30 and emphasizes the motion of Algol 
B between the two epochs. To guide the eye the approximate positions 
of Algol B are indicated at each epoch by a filled black circle. 
The uniformly weighted restoring beam is shown in the lower 
right hand corner. \label{fig1}}
\end{figure}

Initial astrometric results (separation and position angle) were obtained by 
using an image for each night to provide an initial guess for the 
separation and position angle. Algol and $\gamma$ Per were analyzed in the 
same manner. Images of Algol were made using AIPS \citep{aips}, 
DIFMAP and BSMEM \citep{bsmem}. Figure ~\ref{fig1} 
illustrates the motion of Algol B over two epochs and the 
location of Algol C with uniformly weighted images made with DIFMAP. 
This guess was refined in a fit to the visibility data directly. 
As typically done in the reduction of NPOI data, we used a fraction 
of the CLEAN beam (20\% in this case) to provide a more realistic 
estimate of the uncertainty ellipse since the formal errors from the 
fit to the visibility data usually underestimate the true uncertainty 
in the results. These initial astrometric results were used to derive an 
initial fit for the orbital elements. 

For the A$-$B orbit we fixed the eccentricity $e$ 
to zero \citep{soder}. We used the photometric light elements \citep{kim} for the 
A$-$B orbit, and the inclination from the light curve analysis \citep{mtr88}.
A fit was then performed for the position angle of the ascending node $\Omega$, 
and semi-major axis $a$. The Na D lines of Algol B detected by 
\citet{tl} and their orbital elements do verify that we have correctly 
identified the quadrant of the ascending node of Algol A$-$B.

For fitting the AB$-$C orbital elements we began with the orbital elements 
of \citet{pan} and then corrected the position angle of the ascending node. 
We solved for all the orbital elements. Again, comparison with spectroscopic data 
\citep{ebb,hill} verify that the quadrant of the ascending node is 
correctly identified. 

In the final step, these orbital elements were 
refined by comparing them with the visibility data for each scan.
Our final astrometric results appear in  Table~\ref{tbl-1}. 
Differential corrections due to the orbital 
motion of the A$-$B pair are included in Table~\ref{tbl-1} based on the orbital 
elements listed in Table~\ref{tbl-3}. These differential corrections 
were first applied to NPOI data for close binaries as described in 
\citet{hum95}. The relative positions in Table~\ref{tbl-1} 
between composite components in the hierarchical triple, i.e.\ AB-C, 
refer to the photocenter of AB. 

\begin{figure}
\plotone{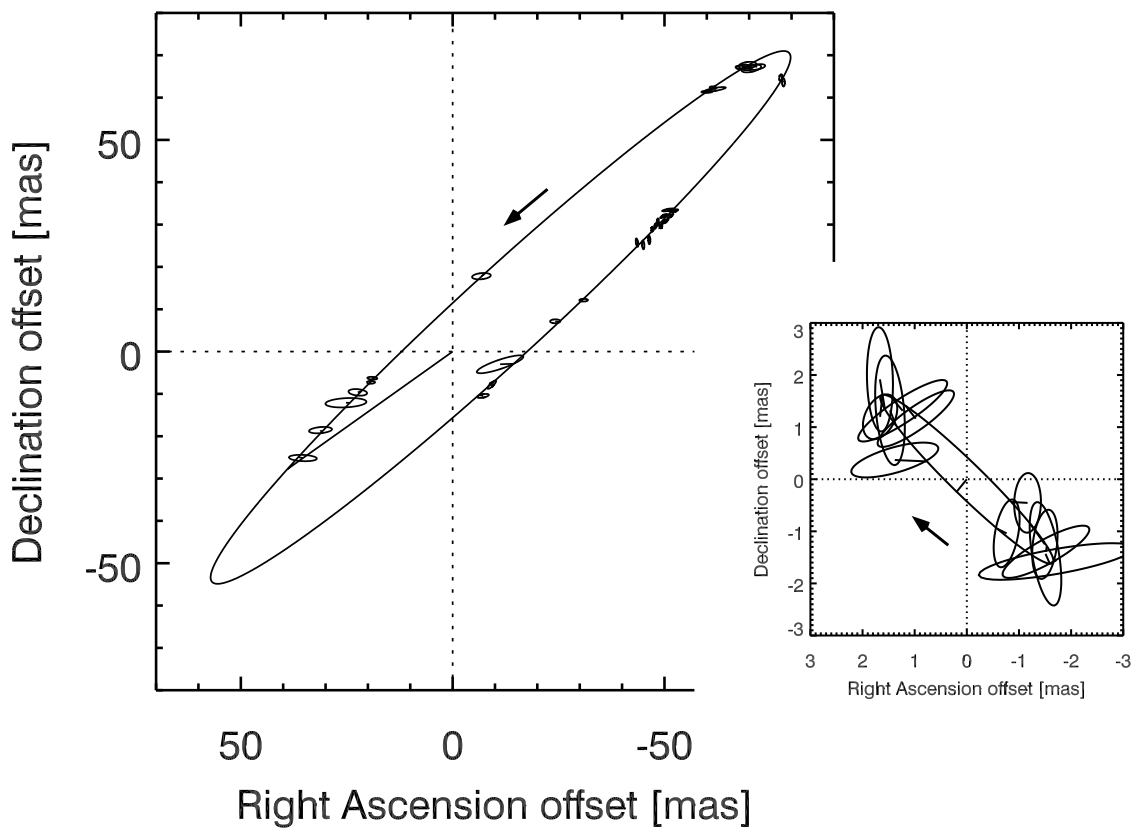}
\caption{The large figure illustrates the AB$-$C orbit. 
A vector from the origin indicates the periastron point.
The inset figure shows the A$-$B orbit with a vector 
from the origin indicating the position of primary 
eclipse minimum light. The astrometric results of Table~\ref{tbl-1} 
are plotted with the astrometry of \citet{pan} rotated by 
180$^{\circ}$ and the orbital elements in
Table~\ref{tbl-3}. Uncertainty ellipses are 20\% of the CLEAN beam for 
the NPOI data. Astrometric positions are fit to the individual 30 second scans. 
The plotted positions for A$-$B are computed at UT07:00 on the date of 
observation between 2006 Oct 19 and 2006 Nov 06. Arrows show the 
direction of the orbital motion on the sky.
\label{fig2}}
\end{figure}

Using the Algol C orbital elements we again solved for the magnitude 
differences (B and C relative to A) and the three elements 
(a, $\Omega$ \& P) of the A$-$B pair. A final fit for the stellar 
masses was performed using the radial velocities contained 
in \citet{tl,hill} and \citet{hill93} for Algol A, B, and C. 
The masses in Table~\ref{tbl-2} served as initial guesses 
for the fits. The results and uncertainties are given in Table~\ref{tbl-3}. 
Figure ~\ref{fig2} shows the orbits of the Algol system with the 
astrometric data.

\section{Discussion}

Our discussion of the analysis of the interferometric observations 
will be restricted to the new insights we have gained on this well
studied system. Our NPOI observations mark the first resolution of the Algol 
system into three components. \citet{csiz} reported the resolution of the 
close binary with the CHARA array at near infrared wavelengths but without 
absolute phase calibration. Our orbital solution is fully consistent 
with the pioneering radio interferometric observations of \citet{les93}. 
We have unambiguously determined that the close pair orbit is retrograde,
and nearly orthogonal to the plane of the wide orbit. The relative
angle $\phi$ is $96^{\circ} \pm 5^{\circ}$ according to 

\[
\cos \phi = \cos i_1\cos i_2+\sin i_1\sin i_2\cos(\Delta \Omega),
\]

\noindent where  $i_1$ and $i_2$ are the inclinations of the two orbits,
and $\Delta\Omega$ is the difference between the two ascending
node angles. This improvement to the orbital plane orientation 
is relevant to dynamical studies of hierarchical triples 
\citep[e.g.][]{kem98}. This orientation of the wide orbit also removes 
the discrepancy expected between the photocenter motion one would compute 
using the AB$-$C elements of \citet{pan} and that found from HIPPARCOS or the 
orbit of \citet{heintz94}. Our retrograde orientation of the close 
binary orbit contrasts with the prograde orbit of \citet{csiz}. 
The position angle calibration of \citet{csiz} depended on 
VLBI observations made during a radio flare of Algol which may 
have complicated their analysis due to the changing radio morphology of 
Algol \citep[\S 4.1 of ][]{csiz}. As the NPOI observations are 
calibrated to produce an absolute position angle as shown 
in \S 2 (with $\gamma$ Per as the position angle calibrator) 
we are confident that the retrograde orbit of the close binary pair is correct.  
The 15 GHz VLBI observations of \citet{peter} also agree with the  
previously determined retrograde orbit.  

The absolute position angle calibration of the NPOI enables a revision of the 
orientation of the Algol C orbit of \citet{pan}. Our results place the 
maximum AB$-$C separation in the same quadrant as found by Heintz 
\citep{heintz94,gjh}. The Hipparcos orbital solution 
\citep[Double and Multiple Systems Annex]{esa97} references \citet{gjh} 
for the quadrants of the longitude of periastron $\omega$ and the position angle 
of the ascending node $\Omega$, and our observations confirm the accuracy of the 
Hipparcos orbital elements. This confirmation is important because it links 
results from both the optical and radio reference frames. A 180$^{\circ}$ reversal 
of the position angle of the ascending node of the AB$-$C orbit presented here 
would create a time variable systematic offset of the photocenter \citep{gjh} 
that could not be reconciled with the HIPPARCOS observations.   

Our interferometric observations have resulted in the first resolved images 
of the triple system and the first directly measured magnitude differences 
for the three stars in Algol. Previous estimates of the magnitude difference
for the close binary were made by modeling photometric and spectroscopic 
data \citep[e.g.][ and references therein]{kim,mtr88}. The 
V band magnitude differences predicted from these models span 
slightly more than one magnitude: 3.72 $\pm$ 0.10 \citep{w72}, 
2.97 $\pm$ 0.31 \citep{soder}, 3.92 $\pm$ 0.88 \citep{mtr88}, 
and 2.71 $\pm$ 0.15 \citep{kim}. Our directly measured V band 
magnitude differences (Table~\ref{tbl-3}) favor magnitude differences 
of less than 3. Our AB$-$C magnitude difference is in excellent 
agreement with the previously determined value using the Mark III 
\citep{pan} and it is also consistent with an early speckle 
interferometry result \citep{lab74}. We extrapolated the magnitude difference 
to the center wavelength of the Hipparcos Hp filter using the stellar atmosphere 
parameters from Table~\ref{tbl-2}, and used the masses given in that table to 
determine a $18.4$ mas amplitude for the motion of the photo center. 
This result is consistent with the Hipparcos orbital solution 
of $19.0\pm0.6$ mas (ESA 1997). The dynamical parallax determined 
from our full fit to the astrometric data and published radial velocities 
is $34.7\pm0.6$, consistent with the Hipparcos value of $35.1\pm0.9$ mas.

Our magnitude differences add 
a directly measured constraint to the results obtained from modeling the 
photometric light curve and spectroscopic data. It may be useful to 
re-examine the modeling of the close binary using magnitude differences 
derived directly from the interferometer measurements. Other bright double 
and multiple stellar systems will yield similar constraints for use with 
spectroscopic and photometric data. 


\acknowledgments

The Navy Prototype Optical Interferometer is a joint project of the 
Naval Research Laboratory and the US Naval Observatory, in
cooperation with Lowell Observatory, and is funded by the Office of
Naval Research and the Oceanographer of the Navy. The authors would
like to thank Jim Benson and the NPOI observational support staff 
whose efforts made this project possible. S. Bosken \& G. Shelton 
of the USNO Library were very helpful in our literature search. 
RTZ is thankful for the support of the Fizeau Exchange Visitor Program 
through the European Interferometry Initiative (EII) and 
OPTICON (an EU funded framework program, contract number RII3-CT-2004-001566). 
This work was partially supported by National Science Foundation grant
AST-0908440 to Richards. The literature search made use of the 
NASA ADS abstract service, the SIMBAD database maintained at CDS, 
Strasbourg, France and the JSTOR archives. We thank an anonymous referee 
for numerous suggestions which improved the paper.

{\it Facilities:} \facility{NPOI ()}

\clearpage




\begin{deluxetable}{ccccc}
\tablecaption{A{\sc dopted} {\sc and} D{\sc erived} S{\sc ystem} P{\sc arameters} \label{tbl-2}}
\tablewidth{0pt}
\tablehead{ \colhead{Parameters} & \multicolumn{3}{c}{Stellar Component} \\
\colhead{} & \colhead{A} & \colhead{B} & \colhead{C} } 
\startdata
Diameter (mas)  & 0.77 & 0.93 & 0.37 \\
Mass (M$_{\odot}$)\tablenotemark{a} & 3.7$\pm$ 0.3 & 0.81$\pm$ 0.05 & 1.6$\pm$ 0.1 \\
$T_{\rm eff}$(K)\tablenotemark{a} & 13000 & 4500 & 7500 \\
log {\it g} & 4.0 & 3.5 & 4.5 \\
\enddata
\tablecomments{Values for diameters calculated as described in \S3. The log(g) 
values are from \citet{mtr93} rounded to match the atmosphere models we used.
Diameter, $T_{\rm eff}$ and log {\it g} were fixed during the model-fitting. 
The masses here were initial estimates in the model fitting and the final 
mass results and uncertainties are listed in Table~\ref{tbl-3}.}
\tablenotetext{a}{\citet{mtr93}}
\end{deluxetable}
\clearpage

\begin{deluxetable}{ccrrccrrrc}
\tabletypesize{\scriptsize}
\rotate
\tablecaption{NPOI R\sc{elative} A\sc{strometric} R\sc{esults}\label{tbl-1}}


\tablehead{
 \colhead{} & \colhead{} & \multicolumn{2}{c}{AB-C} &  \multicolumn{2}{c}{A-B} 
& \multicolumn{3}{c}{Error Ellipse\tablenotemark{a}} \\
\colhead{UT Date} & \colhead{JY}
& \colhead{$\rho$} & \colhead{$\theta$} 
& \colhead{$\rho$} & \colhead{$\theta$} 
& \colhead{$\sigma_{maj}$} & \colhead{$\sigma_{min}$} & \colhead{$\phi$} 
& \colhead{$\Phi$} \\ 
\colhead{} & \colhead{(yrs)} 
& \colhead{(mas)} & \colhead{(deg)} 
& \colhead{(mas)} & \colhead{(deg)} 
& \colhead{(mas)} & \colhead{(mas)} & \colhead{(deg)} 
& \colhead{} } 

\startdata
1997 Oct 16&1997.7901&100.63&309.22
& \nodata & \nodata
&0.996&0.334&  2.0
& \nodata
\\
1997 Oct 17&1997.7928&100.94&309.96
& \nodata & \nodata
&0.790&0.366&168.0
& \nodata
\\
1999 Mar  4&1999.1700& 12.11&228.86
& \nodata & \nodata
&1.360&0.396&133.9
& \nodata
\\
2006 Oct 19&2006.7976& 50.56&300.60
& 2.08&226.49
&1.006&0.256&  9.6
&0.78
\\
2006 Oct 20&2006.8004& 51.56&299.26
& 1.98& 48.10
&1.056&0.270&  5.0
&0.14
\\
2006 Oct 23&2006.8086& 53.38&299.58
& 2.54& 41.01
&1.002&0.248&181.5
&0.18
\\
2006 Oct 27&2006.8195& 55.30&301.52
& 1.25&248.71
&0.572&0.256& -1.4
&0.57
\\
2006 Oct 28&2006.8223& 56.32&301.79
& 1.29&216.29
&0.660&0.222&169.6
&0.93
\\
2006 Oct 29&2006.8250& 57.36&301.01
& 2.05& 54.25
&0.460&0.286&153.2
&0.27
\\
2006 Oct 30&2006.8277& 57.38&302.45
& 1.95&230.14
&0.668&0.232&173.4
&0.62
\\
2006 Oct 31&2006.8305& 58.26&302.82
& \nodata & \nodata
&0.440&0.272&146.0
& \nodata
\\
2006 Nov  1&2006.8332& 58.78&301.49
& 1.80& 43.12
&1.014&0.228&123.5
&0.32
\\
2006 Nov  2&2006.8360& 59.21&302.73
& 2.06&227.45
&0.944&0.232&119.1
&0.67
\\
2006 Nov  3&2006.8387& 60.16&302.22
& \nodata & \nodata
&0.924&0.204&129.3
& \nodata
\\
2006 Nov  4&2006.8414& 60.55&301.92
& 1.52& 40.09
&0.886&0.220&125.4
&0.36
\\
2006 Nov  5&2006.8442& 60.95&303.31
& 2.33&227.33
&1.506&0.238&100.0
&0.72
\\
2006 Nov  6&2006.8469& 62.08&302.41
& 1.43& 75.06
&0.866&0.242&106.1
&0.06
\\
2008 Oct 27&2008.8208& 83.00&306.52
& \nodata & \nodata
&0.564&0.292&11.1
&\nodata
\\ 
\enddata

\tablecomments{Col. 1: UT Date of observation; 
Col. 2: Julian year at 7 hours UT; Cols. 3-6: separation,
position angle (from north through east) for the AB-C and A-B
components, respectively; 
Col. 7: Semi-major axis of error ellipse; Col. 8: Semi-minor axis of error ellipse; 
Col. 9: Position angle of error ellipse;
Col. 10: Close binary orbital phase 
$\Phi$ using light elements $2441773.49 + 2.8673285*E$ \citep{kim}. 
$\Phi$ of 0.0 $\equiv$ primary eclipse, 0.5 $\equiv$ secondary eclipse.}
\tablenotetext{a}{The error ellipse is the uncertainty in the location 
of the position vector. For component C this is with respect 
to the AB photocenter.}

\end{deluxetable}
\clearpage





\begin{deluxetable}{ccccc}
\tablecaption{O{\sc bital} S{\sc olution} A{\sc nd} C{\sc omponent P{\sc arameters}}\label{tbl-3}}
\tablewidth{0pt}
\tablehead{ \colhead{Orbital} & \colhead{A$-$B} & \colhead{A$-$B} & \colhead{AB$-$C} & 
\colhead{AB$-$C} \\
\colhead{Element} & \colhead{\citet{soder}} & \colhead{This work} & \colhead{\citet{pan}} & \colhead{This work}}
\startdata
a (mas)  & 2.2 $\pm$ 0.1 & 2.3 $\pm$ 0.1 & 94.61 $\pm$ 0.22 & 93.8 $\pm$ 0.2\\
i (deg)  & 81.4 $\pm$ 0.2\tablenotemark{a} & 98.6 \tablenotemark{c}& 83.98 $\pm$ 0.09 & 83.7 $\pm$ 0.1 \\
$\Omega$ (deg) & 132 $\pm$ 4 & 47.4 $\pm$ 5.2 & 312.26 $\pm$ 0.13 & 132.7 $\pm$ 0.1 \\
e        & 0 & 0 & 0.225 $\pm$ 0.005 & 0.225 $\pm$ 0.005 \\
$\omega$ (deg)& \nodata & \nodata & 310.29 $\pm$ 0.08 & 310.8 $\pm$ 0.1 \\
T$_0$ (JY) & \nodata & 1973.2471\tablenotemark{b} & 1987.3689 & 1987.3689 \\
T$_0$ (JD) & \nodata & 2441773.49\tablenotemark{b} & 2446931.4 $\pm$ 1.5 & 2446931.6 $\pm$ 0.1 \\
P (days) & 2.8673 & 2.867328 & 680.05 $\pm$ 0.06 & 679.85 $\pm$ 0.04 \\
P (years) & \nodata &  \nodata  & 1.8619 $\pm$ 0.0002 & 1.8613 $\pm$ 0.0001 \\
$\pi_{dyn}$ (mas) & & & & 34.7 $\pm$ 0.6 \\
Magnitude Differences & & & & \\
Components & $\Delta$m(550nm) & $\Delta$m(800nm) & & \\
A$-$B  & 2.70 $\pm$ 0.3 & 2.20 $\pm$ 0.3 \\
A$-$C  & 2.8 $\pm$ 0.2 & 2.6 $\pm$ 0.2 \\ 
AB$-$C & 2.9 $\pm$ 0.1 & 2.7 $\pm$ 0.1 \\
Masses (M$_{\odot}$) & & & & \\
M(A) & 3.7 $\pm$ 0.2 & & & \\ 
M(B) & 0.8 $\pm$ 0.1 & & & \\ 
M(C) & 1.5 $\pm$ 0.1 & & & \\
\enddata

\tablenotetext{a}{\citet{mtr88}}
\tablenotetext{b}{Minimum light of primary eclipse}
\tablenotetext{c}{i $>$ 90$^{\circ}$ used to indicate retrograde motion as defined by \citet{heintz}.}

\end{deluxetable}

\clearpage

\clearpage

\clearpage

\end{document}